\documentclass{article}

\usepackage[latin1]{inputenc}
\usepackage{times}
\usepackage{latexsym}
\usepackage{amsmath}
\usepackage{amssymb}
\usepackage{graphicx}
\usepackage{a4wide}
\usepackage{ifthen}

\newcommand{\nc}{\newcommand}
\nc{\rnc}{\renewcommand}
\nc{\nev}{\newenvironment}

\rnc{\subsection}{\secdef\ssa\ssb}
\nc{\ssa}[2][default]{\par\vspace{2ex}\refstepcounter{subsection}\noindent\textbf{\thesubsection. #1. }}
\nc{\ssb}[1]{\par\bigskip\noindent\textbf{#1. }}

\makeatletter
\rnc{\@seccntformat}[1]{{\normalfont\bfseries{\csname the#1\endcsname}\hspace{1pt}.\hspace{0.4em}}}
\rnc{\section}{\@startsection
        {section}%
        {1}%
        {0mm}%
        {-1.5\baselineskip}%
        {\baselineskip}%
        {\normalfont\normalsize\bfseries\centering}%
}
\renewcommand{\@makecaption}[2]{\begin{center}#1. #2\end{center}}
\makeatother

\newcounter{theo}[section]
\rnc{\thetheo}{\thesection.\arabic{theo}}
 
\nev{defform}[2]{\refstepcounter{theo}\begin{list}{}{%
      \setlength{\labelwidth}{0em}%
      \setlength{\leftmargin}{0em}%
      \setlength{\itemindent}{\labelsep}%
      \setlength{\listparindent}{1.5em}
      \setlength{\parsep}{0em}}%
      \item[\textbf{#2 \thetheo\ifthenelse{\equal{#1}{}}{}{ (#1)}.}]}%
    {\end{list}}

\nev{satzform}[2]{\begin{defform}{#1}{#2}\itshape}{\end{defform}}

\nev{theorem}[1][]{\begin{satzform}{#1}{Theorem}}{\end{satzform}}
\nev{lemma}[1][]{\begin{satzform}{#1}{Lemma}}{\end{satzform}}
\nev{corollary}[1][]{\begin{satzform}{#1}{Corollary}}{\end{satzform}}
\nev{proposition}[1][]{\begin{satzform}{#1}{Proposition}{}}{\end{satzform}}
\nev{fact}[1][]{\begin{satzform}{#1}{Fact}}{\end{satzform}}
\nev{example}[1][]{\begin{defform}{#1}{Example}}{\end{defform}}
\nev{definition}[1][]{\begin{defform}{#1}{Definition}}{\end{defform}}
\nev{remark}[1][]{\begin{defform}{#1}{Remark}}{\end{defform}}

\nc{\proof}{\medskip\noindent\textit{Proof: }}
\nc{\proofend}{\hfill$\Box$\vspace{\topsep}\par}

\nc{\FOR}{\textbf{for}}
\nc{\FORALL}{\textbf{for all}}
\nc{\TO}{\textbf{to}}
\nc{\DO}{\textbf{do}}
\nc{\OD}{\textbf{od}}
\nc{\IF}{\textbf{if}}
\nc{\FI}{\textbf{fi}}
\nc{\THEN}{\textbf{then}}
\nc{\ELSE}{\textbf{else}}
\nc{\WHILE}{\textbf{while}}
\nc{\REPEAT}{\textbf{repeat}}
\nc{\UNTIL}{\textbf{until}}
\nc{\OR}{\textbf{or}}
\nc{\AND}{\textbf{and}}
\nc{\ACCEPT}{\textnormal{ACCEPT}}
\nc{\REJECT}{\textnormal{REJECT}}

\nc{\im}[1]{\item\hspace{#1cm}}
\nev{algorithm}{\begin{enumerate}\rnc{\labelenumi}{\textit{\small \arabic{enumi}.}}\rnc{\itemsep}{0ex}}{\end{enumerate}}

\nc{\prob}[3][9]{\begin{center}\normalfont\fbox{\begin{tabular}[t]{rp{#1cm}}\textit{Input:}&#2.\\\textit{Problem:}&#3.\end{tabular}}\end{center}}

\rnc{\labelenumi}{(\arabic{enumi})}
\rnc{\labelitemi}{\text{--}}
\rnc{\phi}{\varphi}
\nc{\bigmid}{\;\big|\;}
\nc{\Bigmid}{\;\Big|\;}

\nc{\FO}{\textup{FO}}
\nc{\tp}{\textup{tp}}
\nc{\tw}{\textup{tw}}
\nc{\ltw}{\textup{ltw}}
\nc{\diam}{\textup{diam}}
\rnc{\max}{\textup{max}}
\rnc{\min}{\textup{min}}
\nc{\true}{\textup{true}}
\nc{\false}{\textup{false}}
\nc{\val}{\textup{val}}
\nc{\CC}{\textup{C}}

\begin{document}

\title{\large\bf Deciding first-order properties of locally tree-decomposable structures}
\author{\normalsize Markus Frick\hspace{2cm}\normalsize Martin Grohe\\
\small Institut f\"ur Mathematische Logik, Eckerstr.~1, 79104
Freiburg, Germany\\
\small Email: \{frick,grohe\}@logik.mathematik.uni-freiburg.de}
\date{\normalsize\today}
\maketitle

{\rnc{\thefootnote}{}\footnotetext{A preliminary version appeared in
    \emph{Proceedings of the 26th International Colloquium on
      Automata, Languages, and Programming}, Lecture Notes in Computer
    Science 1644, pp.331-340. \copyright\ Springer-Verlag 1999}}

\begin{abstract}
  We introduce the concept of a class of graphs, or more generally,
    relational structures, being \emph{locally
    tree-decom\-posable}. There are numerous examples of locally
  tree-decomposable classes, among them the class of planar graphs and
  all classes of bounded valence or of bounded tree-width. We also
    consider a slightly more general concept of a class of structures
    having \emph{bounded local tree-width}.
  
    We show that for each property $\phi$ of structures that is
    definable in first-order logic and for each locally
    tree-decomposable class $\CC$ of graphs, there is a
    linear time algorithm deciding whether a given structure $\mathcal
    A\in \CC$ has property $\phi$. For classes $\CC$ of
    bounded local tree-width, we show that for every $k\ge 1$ there is an
    algorithm that solves the same problem in time $O(n^{1+(1/k)})$ (where
    $n$ is the cardinality of the input structure).
\end{abstract}

\section{Introduction}
It is an important task in the theory of algorithms to find feasible
instances of otherwise intractable algorithmic problems. A notion that
has turned out to be extremely useful in this context is that of
\emph{tree-width} of a graph. \textsc{3-Colorabil\-ity},
\textsc{Hamiltonicity}, and many other NP-complete properties of
graphs can be decided in linear time when restricted to graphs whose
tree-width is bounded by a fixed constant (see \cite{bod97} for a
survey).

Courcelle \cite{cou90} proved a meta-theorem, which easily implies
numerous results of the abovementioned type: \emph{Let $w\ge 1$ and
  $\phi$ be a property of graphs that is definable in monadic
  second-order logic. Then $\phi$ can be decided in linear time on
  graphs of tree-width at most $w$.} As a matter of fact, this result
does not only hold for graphs, but for arbitrary relational
structures. Although Courcelle's theorem does
not give practical algorithms, because the hidden constants are too
big, it is still useful since it gives a simple way to recognize a
property as being linear time decidable on graphs of bounded
tree-width. Once this has been done, a more refined analysis using the
combinatorics of the particular property may yield a practical
algorithm.

Though maybe the most successful, bounded tree-width is not the
only restriction on graphs that makes algorithmic tasks easier. Other
useful restrictions are \emph{planarity} or \emph{bounded
  valence}.
For example, consider the problem \textsc{$k$-Dominating Set} for a
fixed $k$. (Given a graph $\mathcal G$, is there a set $D$ of at most $k$
vertices of $\mathcal G$ such that every vertex of $\mathcal G$ is either equal or
adjacent to a vertex in $D$?) To solve \textsc{$k$-Dominating Set} in general,
we do not know much better than just trying all $O(n^k)$ candidate
sets ($n$ always denotes the number of vertices of the input
graph). However, on planar graphs \textsc{$k$-Dominating Set} can be
solved in time $O(11^kn)$, and on graphs of valence at most $l$, it
can be solved in time $O((l+1)^kn)$ \cite{dowfel99}.

Unfortunately, the analogue of Courcelle's theorem does not hold for
planar graphs or classes of bounded valence; \textsc{3-Colorability}
is a monadic second-order definable property that remains NP-complete
when restricted to the class of planar graphs of valence at most 4
\cite{garjohsto76}. Instead of monadic second-order, we study the
complexity of first-order definable properties.

Seese was the first to give a meta-theorem in the style of Courcelle's
theorem for a more general class of structures; in \cite{see96} he
proved that for every $l\ge 1$ and for every first-order definable
property of structures there is a linear time algorithm that decides
whether a given structure of valence at most $l$ has this property.

An observation that has been used for various algorithms on planar
graphs (essentially it goes back to Baker \cite{bak94}) is that there
is a bound on the tree-width of a planar graph only depending on its
diameter. A different way to see this is that a local neighborhood of
a vertex in a planar graph has tree-width bounded by a number only
depending on the radius of this neighborhood.  As a matter of fact,
given a planar graph $\mathcal G$ we can compute in linear time a family of
subgraphs of bounded tree-width such that a suitably big neighborhood
of every vertex of $\mathcal G$ is completely contained in one of these
subgraphs.

We call classes of graphs admitting such a covering algorithm
\emph{locally tree-decomposable} (a precise definition is given in
Section \ref{sec:tc}). Examples of locally tree-decomposable classes
of graphs are all classes of bounded genus, bounded valence, and
bounded tree-width. The concept easily generalizes to arbitrary
relational structures.

Eppstein \cite{epp99} considered a closely related, though slightly
weaker concept he called the \emph{diameter-treewidth property} (we
call this property \emph{bounded local tree-width} and refer the
reader to Section \ref{sec:ltw} for the definition). Eppstein proved
that the subgraph isomorphism problem for a fixed subgraph $\mathcal
H$, asking whether a given graph $\mathcal G$ contains $\mathcal H$,
is solvable in linear time when restricted to graphs $\mathcal G$
contained in a class of graphs that is closed under taking minors and
has bounded local tree-width. It is not hard to see that every class
$\CC$ of graphs that is closed under taking minors and has bounded local
tree-width is locally tree-decomposable (cf.\ Lemma \ref{lem:mintc}).

Thus our main result goes much further: 

\begin{theorem}\label{theo:mt1}
Let $\CC$ be a class of relational structures that is locally
tree-decomposable and $\phi$ a property definable in
first-order logic. Then there is a linear time algorithm deciding
whether a given structure $\mathcal A\in \CC$ has property $\phi$.
\end{theorem}

It may be worth mentioning that in the terminology of \cite{var82},
our result can be rephrased as follows: When restricted to a locally
tree-decomposable class of structures, the \emph{data complexity} of
first-order logic is in linear time.

Examples of first-order definable properties are
\textsc{$k$-Dominating-Set} and \textsc{$k$-Indepen\-dent-Set} for a
fixed $k$, \textsc{$\mathcal H$-Subgraph-Isomorphism} (Given $\mathcal
G$, is $\mathcal H\subseteq \mathcal G$?) and \textsc{$\mathcal
  H$-Homo\-morphism} (Given $\mathcal G$, is there a homomorphism
$h:\mathcal H\rightarrow \mathcal G$?) for a fixed $H$,
\textsc{$(\mathcal H,\mathcal K)$-Extension} (Given $\mathcal G$, is
every $\mathcal H\subseteq \mathcal G$ contained in some $\mathcal
K\subseteq \mathcal G$?) for fixed $\mathcal H\subseteq \mathcal K$.
Let us also give a few examples of a problems defined on other
relational structures than graphs. For $k\ge 1$,
\textsc{$k$-Set-Cover} is the problem of deciding whether a given
family $\mathcal F$ of sets has a subfamily $\mathcal S$ of size at
most $k$ such that $\bigcup\mathcal S=\bigcup\mathcal F$. For $d\ge
1$, \textsc{$(k,d)$-Circuit-Satisfiability} is the problem of deciding
whether a given Boolean circuit of depth at most $d$ has a satisfying
assignment in which at most $k$ input gates are set to `\true'. Both
\textsc{$k$-Set-Cover} and \textsc{$(k,d)$-Circuit-Satisfiability} can
be seen as first-order definable problems on certain relational
structures. Thus our theorem implies, for example, that
\textsc{$k$-Set-Cover} can be solved in linear time for set systems
where each element is only contained in a bounded number of sets, and
that \textsc{$(k,d)$-Circuit-Satisfiability} can be solved in linear
time for circuits whose underlying graph is planar.  Of course
problems like \textsc{Subgraph-Isomorphism}, \textsc{Homo\-morphism},
\textsc{Extension} can be generalized arbitrary relational structures.

As a last example, let us consider the problem of evaluating a
(Boolean) database query formulated in the relational calculus against
a relational database. Since relational calculus is the same as
first-order logic, and relational databases are just finite relational
structures, our theorem applies and shows, for example, that
Boolean relational calculus queries can be evaluated in linear time on
databases whose underlying graph is planar. As a matter of fact, this last
example was one of our main motivation for starting this research. It
seems that when storing geographical data such as road maps, planar
structures come up quite naturally.

Thus our theorem gives a unifying framework for various results
solving concrete problems on specific locally tree-decom\-posable
classes such as the class of planar graphs. In addition, it yields a
number of new results of this type.

Using the same techniques, we prove another theorem that applies to the
even more general context of classes of structures of bounded local
tree-width:

\begin{theorem}\label{theo:mt2}
  Let $\CC$ be a class of relational structures of bounded local
  tree-width and $\phi$ a first-order definable property. Then for
  every $k\ge 1$ there is an algorithm deciding whether a given
  structure $\mathcal A\in \CC$ has property $\phi$ in time $O(n^{1+(1/k)})$.
\end{theorem}

The complexity of first-order properties of relational structures has
been studied under various aspects. It is long known that every
first-order property of graphs can be decided in polynomial time,
actually in $\textrm{AC}_0$ \cite{ahoull79,imm82}. A question closer
to our theorem is whether deciding first-order properties is
\emph{fixed-parameter tractable}, that is, whether there is a fixed
$c$ such that every first-order property of finite relational
structures can be decided in time $O(n^c)$. This question has been
brought up by Yannakakis \cite{yan95}. The theory of fixed-parameter
tractability gives some evidence that the answer is no, as has been
independently proved by Downey, Fellows, Taylor \cite{dowfeltay96} and
Papadimitriou, Yannakakis \cite{papyan97} (deciding first-order
properties is $\textrm{AW}[1]$-complete). Theorem \ref{theo:mt2} shows
that deciding first-order properties of structures in a class of
bounded local tree-width is fixed-parameter tractable. Furthermore, it
has been used in \cite{flugro99} to show that for every class $\CC$ of
graphs such that there is some graph that is not a minor of any graph
in $\CC$, deciding first-order properties of graphs in $\CC$ is
fixed-parameter tractable.

The proofs of our results combine three main ingredients: a
refinement of Courcelle's Theorem \cite{cou90} mentioned above,
Gaifman's Theorem \cite{gai81} stating that first-order properties are
local, and algorithmic techniques based on ideas of Baker \cite{bak94}
and Eppstein \cite{epp99}. To prove Theorem \ref{theo:mt2}, we also
use covering techniques due to Awerbuch and Peleg \cite{awepel90,pel93}.

\section{Preliminaries}

A \emph{vocabulary} is a finite set of relation symbols. Associated
with every relation symbol $R$ is a positive integer called the
\emph{arity} of $R$. In the following, $E$ always denotes a binary relation
symbol and $\tau$ a vocabulary.

A {\em $\tau$-structure} $\mathcal A$ consists of a non-empty set $A$, called
the {\em universe} of $\mathcal A$, and a relation $R^{\mathcal
  A}\subseteq A^r$ for each $r$-ary relation symbol $R\in\tau$. 
If $\mathcal A$ is a $\tau$-structure and $B\subseteq A$, then
$\langle B\rangle^{\mathcal A}$ denotes the substructure induced by
$\mathcal A$ on $B$, that is, the $\tau$-structure $\mathcal B$ with
universe $B$ and $R^{\mathcal B}:=R^{\mathcal A}\cap B^r$ for every
$r$-ary $R\in\tau$.

For instance, we consider \emph{graphs} as $\{E\}$-structures $\mathcal
G=(G,E^{\mathcal G})$, where the binary relation $E^{\mathcal G}$ is
symmetric and anti-reflexive (i.e.\ graphs are undirected and
loop-free). As another example, we can view hypergraphs as
$\{E,P\}$-structures, where $E$ is binary and $P$ unary. A hypergraph
with vertices $V$ and hyperedges $\mathcal H\subseteq\text{Pow}(V)$ is
modeled by the $\{E,P\}$-structure $\big(V\cup\mathcal H,\{(v,H)\mid v\in H\},V\big)$.

\emph{In this paper we only consider finite structures.} Let us remark
that all the results of this paper remain true if we also admit
constants in our structures. We restrict our attention to the
relational case because constants would not give us additional
insights.

\medskip
The formulas of \emph{first-order logic} \FO\ are build up in the usual way
from an infinite supply of variables denoted by $x,y,x_1,\ldots$, the
equality symbol $=$ and relation symbols of a vocabulary $\tau$, the connectives
$\wedge,\vee,\neg,\rightarrow$, and the quantifiers $\forall,\exists$
ranging over the universe of the structure. For example, the
first-order sentence 
\[
\phi:=\forall x_1\forall x_2\forall x_3\big((Ex_1x_2\wedge Ex_1x_3\wedge
Ex_2x_3)\rightarrow\exists y(Ex_1y\wedge Ex_2y\wedge Ex_3y)\big)
\]
says that every triangle of a graph is contained in a $K_4$ (a
complete graph on four vertices). The formula
\[
Px\wedge\neg\exists y\exists z(\neg y= z\wedge Exy\wedge Exz)
\]
defines the set of all vertices $x$ of a hypergraph that are contained in
at most one hyperedge.

A \emph{free variable} in a
first-order formula is a variable $x$ not in the scope of a quantifier
$\exists x$ or $\forall x$. A \emph{sentence} is a formula without free
variables.  The notation $\phi(x_1,\ldots,x_k)$ indicates that all
free variables of the formula $\phi$ are among $x_1,\ldots,x_k$; it
does not necessarily mean that the variables $x_1,\ldots,x_k$ all
appear in $\phi$.  For a
formula $\phi(x_1,\ldots,x_k)$, a structure $\mathcal A$, and
$a_1,\ldots,a_k\in A$ we write $\mathcal A\models
\phi(a_1,\ldots,a_k)$ to say that $\mathcal A$
satisfies $\phi$ if the variables $x_1,\ldots,x_k$ are interpreted by
the vertices $a_1,\ldots,a_k$, respectively.  

\begin{example}
In this example we show how to model the \textsc{$k$-Set-Cover}
problem mentioned in the introduction by a first-order definable
problem. We can view a family $\mathcal F$ of a sets as the hypergraph
whith vertex set $\bigcup\mathcal F$ and edge set $\mathcal F$. 

Let 
\[
\phi_k:=\exists x_1\ldots x_k\forall
y\Big(Py\to\big(Eyx_1\vee\ldots\vee Eyx_k\big)\Big).
\]
Then the hypergraph corresponding to the family $\mathcal F$ satisfies
$\phi_k$ if and only if there exists an $\mathcal S\subseteq\mathcal
F$ of cardinality $|\mathcal S|=k$ such that $\bigcup\mathcal
S=\bigcup\mathcal F$.
\end{example}

\medskip
We often denote tuples $(a_1,\ldots,a_k)$ of elements
of a set $A$ by $\bar a$, and we write $\bar a\in A$ instead of $\bar
a\in A^k$. Similarly, we denote tuples of variables by
$\bar x$.

\medskip 
Our underlying model of computation is the standard RAM-model with
addition and subtraction as arithmetic operations (cf.\ 
\cite{ahohopull74, emd90}). In our complexity analysis we use the
uniform cost measure. 
Structures are represented on a RAM in a straightforward way by
listing all elements of the universe and then all tuples in the
relations. For details we refer the reader to \cite{flufrigro00}.
We define the \emph{size} of a $\tau$-structure $\mathcal A$ to be
$||\mathcal A||:=|A|+\sum_{R\in\tau\;r\text{-ary}}r\cdot|R^{\mathcal
  A}|$; this is the length of a reasonable representation of $\mathcal A$
(if we suppress details that are inessential for us).

\section{Gaifman's Theorem}
The \emph{Gaifman graph} of a $\tau$-structure $\mathcal
A$ is the graph $\mathcal G_{\mathcal A}$ with vertex set $G_{\mathcal
  A}:=A$ and an edge between two vertices $a,b\in A$ if there exists
an $R\in\tau$ and a tuple $(a_1,\ldots,a_k)\in R^{\mathcal A}$ such
that $a,b\in\{a_1,\ldots,a_k\}$.  The \emph{distance} $d^{\mathcal
  A}(a,b)$ between two elements $a,b\in A$ of a structure $\mathcal A$
is the length of the shortest path in $\mathcal G_{\mathcal A}$
connecting $a$ and $b$. For $r \ge 1$ and $a \in A$ we define the
\emph{$r$-neighborhood} of $a$ in $\mathcal A$ to be $N^{\mathcal
  A}_r(a) := \{ b \in A\mid d^{\mathcal A}(a,b) \le r \}$. For a
subset $B\subseteq A$ we let $N_r^{\mathcal A}(B):=\bigcup_{b\in
  B}N_r^{\mathcal A}(b)$.

For every $r\ge 0$ there is a first-order formula $\delta_r(x,y)$ such
that for all $\tau$-structures $\mathcal A$ and $a,b\in A$ we have $\mathcal
A\models\delta_r(a,b)\iff d^{\mathcal A}(a,b)\le r$. For example, if
$\tau=\{E,T\}$ consists of a binary and a ternary relation symbol, we let
\begin{align*}
\delta_0(x,y):=&(x=y)\\
\delta_1(x,y):=&\delta_0(x,y)\vee Exy\vee Eyx\vee\exists z\big(Txyz\vee
Tyxz\vee Txzy\vee Tyzx\vee Tzxy\vee Tzyx\big)\\
\delta_2(x,y):=&\delta_0(x,y)\vee\delta_1(x,y)\vee\exists z\big(\delta_1(x,z)\wedge\delta_1(z,y)\big)
\end{align*}
In the following, we write $d(x,y)\le r$ instead of $\delta_r(x,y)$
and $d(x,y)> r$ instead of $\neg\delta_r(x,y)$.

If $\phi(x)$ is a first-order formula, then $\phi^{N_r(x)}(x)$ is the
formula obtained from $\phi(x)$ by relativizing all quantifiers to
$N_r(x)$, that is, by replacing every subformula of the form $\exists
y\psi(x,y,\bar z)$ by $\exists y(d(x,y)\le r\wedge\psi(x,y,\bar z))$
and every subformula of the form $\forall y\psi(x,y,\bar z)$ by
$\forall y(d(x,y)\le r\rightarrow\psi(x,y,\bar z))$. A formula
$\psi(x)$ of the form $\phi^{N_r(x)}(x)$, for some $\phi(x)$, is
called \emph{$r$-local}.  The basic property of $r$-local formulas
$\psi(x)$ is that it only depends on the $r$-neighborhood of $x$
whether they hold at $x$ or not, that is, for all structures $\mathcal
A$ and $a\in A$ we have $\mathcal A\models\psi(a)\iff\big\langle
N_r^{\mathcal A}(a)\big\rangle\models\psi(a)$.

\begin{theorem}[Gaifman \cite{gai81}]\label{gaifman}
Every first-order sentence is equivalent to a Boolean combination of
sentences of the form
\[
\exists x_1\ldots\exists x_k\big(\bigwedge_{1\le i<j\le
  k}d(x_i,x_j)>2r\wedge \bigwedge_{1\le i\le
  k}\psi(x_i)\big),
\]
for suitable $r,k\ge 1$ and an $r$-local $\psi(x)$.
\end{theorem}

\section{Tree-width}\label{seccou}
A \emph{tree} is an acyclic graph.
A \emph{tree-decomposition} of a $\tau$-structure ${\mathcal A}$ is a
pair $(\mathcal T,(B_t)_{t\in T})$, where $\mathcal T$ is a tree and
$(B_t)_{t\in T}$ a family of subsets of $A$ (called the \emph{blocks}
of the decomposition) such that
\begin{enumerate}
\item For every $a\in A$, the set $\{t\in T\mid a\in B_t \}$
  is non-empty and connected in $\mathcal T$ (that is, induces a subtree).
\item For every $R\in\tau$ and all $\bar a\in R^{\mathcal A}$ there
  is a $t\in T$ such that $\bar a\in B_t$.
\end{enumerate}
The \emph{width} of a tree-decomposition $(\mathcal T,(B_t)_{t\in T})$
is $\max\{|B_t|\mid t\in T \}-1$. The \emph{tree-width} $\tw({\mathcal
  A})$ of ${\mathcal A}$ is the minimal width of a tree-decomposition
of ${\mathcal A}$.

We occasionally use the following simple fact (cf.\ \cite{GMXIII}).

\begin{lemma}\label{lem:linsize}
  Let $w\ge 1$ and $\tau$ a vocabulary. Then there is a constant $c$
  such that for every $\tau$-structure $\mathcal A$ of tree-width at
  most $w$ we have $||\mathcal A||\le c|A|$.
\end{lemma}

Bodlaender \cite{bod96} proved that for each $w\ge 1$ there is a
linear time algorithm that, given a graph $\mathcal G$, either
computes a tree-decomposition of $\mathcal G$ of width at most $w$, or
rejects $\mathcal G$ if $\tw(\mathcal G)>w$. This result is underlying
most of the linear time algorithms on graphs of bounded tree-width.
Using the well-known fact that a structure $\mathcal
A$ has the same tree-width as its Gaifman graph $\mathcal G_{\mathcal
  A}$, Bodlaender's result can easily be extended to arbitrary
relational structures.

Recall Courcelle's theorem that we mentioned in the introduction: 

\begin{theorem}[Courcelle \cite{cou90}]\label{theo:cou}
  Let $w\ge 1$. Then for every sentence $\phi$ of monadic second-order
  logic there is a linear time algorithm that decides whether a given
  structure $\mathcal A$ of tree-width at most $w$ satisfies $\phi$.
\end{theorem}

Monadic second-order logic is an extension of first-order logic that
also allows quantification over sets.

Using known techniques for algorithms on graphs of bounded tree-width,
it is not hard to prove the following lemma (see \cite{flufrigro00}).
We are only going to use the first-order version of the lemma later.

\begin{lemma}\label{lem:cou}
  Let $w\ge 1$. Then for every formula $\phi(x)$ of monadic
  second-order logic there is a linear time algorithm that, given a
  graph $G$ of tree-width at most $w$, computes the set $\phi(\mathcal A):=\{a\in
  V^G\mid G\models\phi(a)\}$.
\end{lemma}

\section{Local Tree-Width}\label{sec:ltw}

\begin{definition}
\begin{enumerate}
\item
The \emph{local tree-width} of a structure ${\mathcal A}$ is the function $\ltw^{\mathcal A}:\mathbb N\rightarrow\mathbb N$ defined by
\[
\ltw^{\mathcal A}(r):=\max\big\{\tw(\langle N^{\mathcal
  A}_r(a)\rangle)\bigmid a\in A\big\}.
\]

\item
A class $\CC$ of structures has \emph{bounded local tree-width} if
  there is a function $f:\mathbb N\rightarrow\mathbb N$ such that
  $\ltw^{{\mathcal A}}(r)\le f(r)$ for all ${\mathcal A}\in\mathcal \CC$, $r\in\mathbb N$.
\end{enumerate}
\end{definition}

\begin{example}\emph{Structures of bounded tree-width.}
  Let ${\mathcal A}$ be a structure of tree-width at most $k$. Then
  $\ltw^{\mathcal A}(r)\le k$ for all $r\in\mathbb N$.
\end{example}

The \emph{valence} of a structure $\mathcal A$ is the maximal number
of neighbors of a vertex $a\in A$ in the Gaifman graph $\mathcal
G_{\mathcal A}$, i.e.\ $\max_{a\in A}|\{b\mid(a,b)\in E^{\mathcal
  G_{\mathcal A}}\}|$.

\begin{example}\label{ex:val}\emph{Structures of bounded valence.}
  Let ${\mathcal A}$ be a structure of valence at most $l$, for an
  $l\ge 1$. Then $\ltw^{\mathcal A}(r)\le l(l-1)^{r-1}$ for all
  $r\in\mathbb N$.
\end{example}

\begin{example}[Robertson and Seymour \cite{GMIII}]\label{ex:planar}\emph{Planar Graphs.}
The class of planar graphs has bounded local tree-width. More
precisely, for every planar graph $G$ and $r\ge 1$ we have
$\ltw^G(r)\le 3r$.
\end{example}

\begin{example}[Eppstein \cite{epp99}]\label{ex:genus}\emph{Graphs of bounded genus.}
Let $S$ be a surface. Then the class of all graphs embeddable in $S$
has bounded local tree-width. More precisely, there is a constant $c$
such that for all graphs $G$ embeddable in $S$ and for all $r\ge 0$ we
have $\ltw^G(r)\le c\cdot g(S)\cdot r$. 
\end{example}

\begin{example}
We can view a simplicial complex as a hypergraph whose vertices are
the corners of the complex. Then it is easy to see that the class of
all simplicial complexes homeomorphic to a 2-manifold has bounded local
tree-width.
\end{example} 

Recall that a \emph{minor} of a graph $\mathcal G$ is a graph
$\mathcal H$ that is obtained from a subgraph of $\mathcal G$ by
contracting edges. The class of planar graphs, and, more generally,
the classes of graphs of bounded genus are examples of classes of
graphs that are closed under taking minors. Eppstein gave
the following nice characterization of all classes of graphs of
bounded loocal tree-width that are closed under taking minors.
An \emph{apex graph} is a graph $G$ that has a
vertex $v\in V^G$ such that $G\setminus\{v\}$ is planar.

\begin{theorem}[Eppstein \cite{epp99a,epp99}]\label{theo:apex}
  Let $\CC$ be a minor-closed class of graphs. Then $\CC$ has bounded local tree-width if, and only if, $\CC$ does
  not contain all apex graphs.
\end{theorem}

This yields further examples of classes of graphs of bounded local
tree-width. For example, for every $n\ge 1$, the class of all graphs
that do not contain the graph $K_{3,n}$ as a minor has bounded local tree-width. ($K_{m,n}$
denotes the complete bipartite graph with parts of size $m$ and $n$,
respectively.)

\medskip
Note that a structure has the same local tree-width as its Gaifman
graph, so Examples \ref{ex:planar} and \ref{ex:genus} and Theorem
\ref{theo:apex} also give rise to examples of classes of structures of
arbitrary vocabularies that have bounded local tree-width.

\medskip
One of the nice things about bounded local tree-width is that the
notion is quite flexible. Think of a structure modeling a subway
map. The Gaifman graph of this structure will probably be close to
planar, but there may be some edges crossing. Therefore, it may be the
case that planar graph algorithms do not apply, although the graph is
almost planar. On the other hand, the local tree-width of the graph is
probably very close to that of a planar graph, and we can still use our
algorithms for graphs of bounded local tree-width.

\section{Neighborhood and tree covers}\label{sec:tc}
To explore the local tree-likeness of structures of bounded local
tree-width we need to cover them by structures of small tree-width in
a suitable way. The most general approach is to use \emph{sparse
  neighborhood covers}, as they have been studied, for instance, in
\cite{awepel90,awebercowpel93,pel93}. 

\begin{definition}
Let $r,s\ge 0$. An \emph{$(r,s)$-neighborhood cover} of a structure $\mathcal
A$ is a family $\mathcal N$ of subsets of $A$ with the following
properties:
\begin{enumerate}
\item
For every $a\in A$ there exists a $N\in\mathcal N$ such that
$N_r^{\mathcal A}(a)\subseteq N$.
\item
For every $N\in\mathcal N$ there exists an $a\in A$ such that
$N\subseteq N_s^{\mathcal A}(a)$.
\end{enumerate}
\end{definition}

We define the \emph{size} of a family $\mathcal N$ of sets to be
$||\mathcal N||:=\sum_{N\in\mathcal N}|N|$. Recall that the size of a
$\tau$-structure $\mathcal A$ is $||\mathcal
A||=|A|+\sum_{R\in\tau\;r-\text{ary}}r|R^{\mathcal A}|$. The algorithm of the
following lemma is an adaptation of an algorithm due to Peleg \cite{pel93} to
our situation. We think it is worthwhile to present our version of
the algorithm in some detail.

\begin{lemma}[Peleg \cite{pel93}]\label{lem:pel}
Let $k\ge 1$. Then there is an algorithm
that, given a graph $\mathcal G$ and an $r\ge 1$, computes an
$(r,2kr)$-neighborhood cover $\mathcal N$ of $\mathcal G$ of size
$||\mathcal N||=O(|G|^{1+(1/k)})$ in time $O\big(\sum_{N\in\mathcal N}||\langle
N\rangle^{\mathcal G}||\big)$.
\end{lemma}

\proof
The algorithm is described in Figure \ref{fig:alg1}. It iteratively
computes a neighborhood cover $\mathcal N$, maintaining a set $H$ of
vertices whose $r$-neighborhood has not yet been covered by a set in
$\mathcal N$. In each iteration step of the main loop in Lines 3--13,
the algorithm picks an arbitrary vertex $a\in H$ and starts to compute
increasing neighborhoods of $a$ (in Lines 6--10) until a certain threshold is reached
(cf.\ Line 10). Then it adds the computed set $N$ to the cover
$\mathcal N$ and removes all points whose neighborhood has now been
covered from $H$, before it goes to the next iteration of the main
loop. This process is repeated until $H$ is empty.
\begin{figure}[ht]

\begin{center}
\fbox{\parbox{10cm}{
\textbf{Input:} Graph $\mathcal G$, $r\ge 1$

\begin{algorithm}
\im{0}
$H:=G$
\im{0}
$\mathcal N:=\emptyset$
\im{0}
\WHILE\ $H\neq\emptyset$ \DO
\im{1}
choose arbitrary $a\in H$
\im{1}
$N:=\{a\}$
\im{1}
\DO
\im{2}
$M:=N$
\im{2}
$L:=N_r^{\mathcal G}(M)\cap H$
\im{2}
$N:=N_r^{\mathcal G}(L)$
\im{1}
\WHILE\ $|N|>n^{1/k}|M|$ \OD
\im{1}
$\mathcal N:=\mathcal N\cup\{N\}$
\im{1}
$H:=H\setminus L$
\im{0}
\OD
\end{algorithm}

\textbf{Output:} $\mathcal N$
}}
\end{center}
\caption{}\label{fig:alg1}
\end{figure}

Now let $\mathcal G$ be a graph, $n:=|G|$, and $r\ge 1$. Let $\mathcal
N$ be the cover computed by the algorithm.

\medskip
\textit{Claim 1. }
For every $a\in G$ there exists a $N\in\mathcal N$ such that
$N_r(a)\subseteq N$.

\smallskip
\textit{Proof: }An element $a$ is removed from the set $H$ of
uncovered elements in Line 12 if it belongs to a set $L$ such that
$N=N_r^{\mathcal G}(L)$ has been added to $\mathcal N$. Of course this
$N$ contains $N_r^{\mathcal G}(a)$. 
This proves Claim 1. 

\medskip
\textit{Claim 2. }
For every $N\in\mathcal N$ there exists an $a\in G$ such that
$N\subseteq N_{2kr}^{\mathcal G}(a)$.

\smallskip
\textit{Proof: }We consider the iteration of the main loop that leads
to the definition of $N$. Let $a$ be the element chosen in Line 4, and
let $N_0:=\{a\}$. Let $l\ge 1$ be the number of times the loop in
Lines 6--10 is repeated. For $1\le i\le l$, let $N_i$ be the value of
$N$ after the $i$th iteration. Then for $1\le i\le l-1$ we have
$|N_i|>n^{1/k}|N_{i-1}|$, and therefore $|N_i|>n^{i/k}$. Thus $l\le
k$.

Furthermore, it is easy to see that for $1\le i\le l$ we have
$N_i\subseteq N_{2ir}^{\mathcal G}(a)$. This implies Claim 2.

\medskip
Claims 1 and 2 show that $\mathcal N$ is indeed an
$(r,2kr)$-neighborhood cover of $\mathcal G$. The following Claim 3
shows that the cover is not too large.

\medskip
\textit{Claim 3. }
$||\mathcal N||\le n^{1+(1/k)}$.

\smallskip
\textit{Proof: }
For $N\in\mathcal N$, and let $M$ be the corresponding set that is
computed in the last iteration of the loop in Lines 6--10 that let to
$N$ (i.e.\ $M$ is the value of $N$ after the second but last
iteration of the loop). 

We first show that for distinct $N_1,N_2\in\mathcal N$ we have
$M_1\cap M_2=\emptyset$. To see this, suppose that $N_1$ is computed
first. Let $H_1$ be the value of $H$ after the iteration of the main
loop in which $N_1$ has been computed. Note that for every $a\in M_1$ and
$b\in H_1$ we have $d^{\mathcal G}(a,b)>r$. Moreover, $M_2\subseteq
N_2\subseteq N_r^{\mathcal G}(H_1)$. Thus $M_1\cap M_2=\emptyset$.

Noting that by the condition of Line 10, for all $N\in\mathcal N$ we
have $|N|\le n^{1/k}|M|$, we obtain
\[
||\mathcal N||=\sum_{N\in\mathcal N}|N|\le n^{1/k}\sum_{N\in\mathcal
  N}|M|\le n^{1/k}\cdot n.
\]
The last inequality holds because the $M$ are disjoint subsets of
$G$. This proves Claim 3.

\medskip
It remains to estimate the running time of the algorithm. We claim
that each iteration of the main loop requires time $O(\langle
N\rangle^{\mathcal G})$, for the $N$ added to $\mathcal N$ in
this iteration. To see this, note that essentially we have to do a
breadth-first search on $N$ starting in $a$. To compute $L$ in Line 8, we may have to explore some
edges not contained in $\langle L\rangle$. However, all these edges belong to
$\langle N_r^{\mathcal G}(L)\rangle=\langle N\rangle$.

It may seem that to check the condition of Line 10 we need
multiplication, which is not available as basic operation of a
standard RAM. However, before we start the main computation we can
produce tables that store the values $m^l$ and $m^l\cdot n$ for $1\le
l\le k$, $1\le m\le n$ in linear time on a standard RAM. (We use the
fact that
\[
(m+1)^l=\sum_{(\epsilon_1,\ldots,\epsilon_l)\in\{0,1\}^l}m^{\sum_{i=1}^l\epsilon_i}
\]
to inductively compute the tables. Remember that we treat $k$ as a
constant.) Then we can use these tables to check the condition of 
Line 10 in constant time.  
\proofend

\begin{corollary}\label{cor:ncov}
  Let $k,r\ge 1$, $\tau$ a vocabulary, and $\CC$ a class of $\tau$-structures of
  bounded local tree-width. Then there is an algorithm that, given a structure
  $\mathcal A\in \CC$, computes an $(r,2kr)$-neighborhood cover $\mathcal N$ of
  $\mathcal A$ of size $||\mathcal N||=O(|A|^{1+(1/k)})$ in time
  $O(|A|^{1+(1/k)})$.
\end{corollary}

\proof Since neighborhoods of radius $2kr$ in structures in $\CC$ have bounded tree-width, by Lemma \ref{lem:linsize} there is a constant
$c$ such that for every structure $\mathcal A\in \CC$, every
$(r,2kr)$-neighborhood cover $\mathcal N$ of $\mathcal A$, and every
$N\in\mathcal N$ we have 
\begin{equation}\label{eq:nc1}
||\langle N\rangle^{\mathcal A}||\le c|N|.
\end{equation}
This implies $||\mathcal A||\le c||\mathcal N||$.

Our algorithm first computes the Gaifman graph $\mathcal G_{\mathcal
  A}$ of the input structure $\mathcal A$, which is possible in time
$O(||\mathcal A||)$. Then it computes an an $(r,2kr)$-neighborhood
cover $\mathcal N$ of $\mathcal A$ of size $||\mathcal
N||=O(|A|^{1+(1/k)})$. By Lemma \ref{lem:pel} and \eqref{eq:nc1}, this is
possible in time $O(||\mathcal N||)=O(|A|^{1+(1/k)})$.
\proofend

The following consequence of the proof of the previous corollary is
worth being noted:

\begin{corollary}\label{cor:avdegbltw}
Let $\tau$ be a vocabulary and $\CC$ be a class of $\tau$-structures of
bounded local tree-width. Then for every $k\ge1$ there is a constant
$c$ such that for all structures $\mathcal A\in \CC$ we have $||\mathcal
A||\le c|A|^{1+(1/k)}$.
\end{corollary}

As a matter of fact, a neighborhood cover is more than we need. Often, the
following weaker notion of a \emph{tree cover} leads to better results.

\begin{definition}
Let $r,w\ge 0$. An \emph{$(r,w)$-tree cover} of a structure $\mathcal
A$ is a family $\mathcal T$ of subsets of $A$ with the following
properties:
\begin{enumerate}
\item
For every $a\in A$ there exists a $T\in\mathcal T$ such that
$N_r^{\mathcal A}(a)\subseteq T$.
\item
For every $T\in\mathcal T$ we have $\tw(\langle T\rangle^{\mathcal A})\le w$.
\end{enumerate}
\end{definition}

Note that an $(r,s)$-neighborhood cover of a structure $\mathcal A$ is an
$(r,\ltw^{\mathcal A}(s))$-tree cover of $\mathcal A$. The following lemma is
implicit in \cite{epp99}:

\begin{lemma}[Eppstein \cite{epp99}]\label{lem:mintc}
  Let $r\ge 0$ and $\CC$ be a class of graphs that is closed under taking minors
  and has bounded local tree-width. Let $f:\mathbb N\rightarrow\mathbb N$ be a
  function bounding the local tree-width of the graphs in $\CC$.

Then there is an algorithm that, given a graph $\mathcal G\in \CC$, computes an $(r,f(2r+1))$-tree cover $\mathcal T$ of $\mathcal G$ of
size $||\mathcal T||=O(|G|)$ in time $O(|G|)$.
\end{lemma}

\proof
Let $\mathcal G\in \CC$ and choose an arbitrary vertex $a_0\in G$. For
$0\le i\le j$, let $G[i,j]:=\{a\in G\mid i\le
d^G(a_0,a)\le j\}$. 

We claim that $\tw(\langle G[i,j]\rangle)\le
f(j-i+1)$. This is immediate if $i=0$ or $i=1$, because then
$G[i,j]\subseteq N_{j}^{\mathcal G}(a_0)$. If $i>1$, we simply contract the
connected subgraph $\langle G[0,i-1]\rangle^{\mathcal G}$ to a single vertex $b_0$. We obtain a minor
$\mathcal G'$ of $\mathcal G$, which is also an element of $\CC$ by our
assumption that $\CC$ is closed under taking minors. $\mathcal G'$ still
contains the set $G[i,j]$ as it is, but this set is contained in
$N_{j-i+1}^{\mathcal G'}(b_0)$. This proves the claim.

The claim implies that for all $r\ge 0$, the family $\mathcal
T:=\{G[i,i+2r]\mid i\ge 0\}$ is an $(r,f(2r+1))$-tree cover of
$\mathcal G$ of size at most $(2r+1)|G|$.
On input $\mathcal G$, we can choose an arbitrary $a_0$ and then compute this
tree cover in linear time by breadth-first search.
\proofend 

The existence of a tree-cover of size linear in the size of the structure and
a linear time algorithm computing such a cover is exactly what we need in our
algorithms of the next section. This justifies the following definition:

\begin{definition}
A class $\CC$ of graphs is \emph{locally tree-decomposable} if there is a
function $g:\mathbb N\to\mathbb N$ and an algorithm that, given a structure
$\mathcal A\in \CC$ and an $r\in\mathbb N$, computes an $(r,g(r))$-tree cover of
$\mathcal A$ of size $O(|A|)$ in time $O(|A|)$.\footnote{The hidden constants
  in the $O(\cdot)$-notation may depend on $r$.}
\end{definition}

\begin{satzform}{}{Examples}
All examples of classes of structures of bounded local tree-width that we gave
in Section \ref{sec:ltw} are actually locally tree-decomposable. 

For Example \ref{ex:val}, classes of structures of bounded valence,
this is trivial: If $\mathcal A$ is a structure of valence $l$ and
$r\ge 0$, then the family $\{N_r^{\mathcal A}(a)\mid a\in A\}$ is an
$(r,l(l-1)^{r-1})$-tree cover of $\mathcal A$.

For all other examples, it follows from Lemma \ref{lem:mintc}.
\end{satzform}

The following proposition is an immediate consequence of the
definition of locally tree-decomposable classes of structures:

\begin{proposition}\label{prop:avdegltd}
Let $\tau$ be a vocabulary and $\CC$ be a locally tree-decomposable class of $\tau$-structures. Then there is a constant
$c$ such that for all structures $\mathcal A\in \CC$ we have $||\mathcal
A||\le c|A|$.
\end{proposition}

We close this section with an example showing that the analogue of
Proposition \ref{prop:avdegltd} for classes of bounded local
tree-width is wrong. Remember Corollary \ref{cor:avdegbltw}, though.

\begin{example}\label{ex:bltwsize}
  We construct a class $\CC$ of graphs of bounded local tree-width such
  that for every constant $c$ there is a graph $\mathcal G\in \CC$ with
  $||\mathcal G||\ge c|G|$.
  
  We use the following theorem due to Erd\"os \cite{erd59}: \emph{For
    all $g,k\ge 1$ there exists a graph of girth greater than $g$ and
    chromatic number greater than $k$.} Remember that the \emph{girth}
  $g(\mathcal G)$ of a graph $\mathcal G$ is the length of the
  shortest cycle in $\mathcal G$ and the chromatic number
  $\chi(\mathcal G)$ of $\mathcal G$ is the least number of colors
  needed to color the vertices of $\mathcal G$ in such a way that no
  two adjacent vertices have the same color. It is easy to see that every
  graph $\mathcal G$ with $\chi(\mathcal G)\ge k$ has a connected subgraph
  $\mathcal H$ with average degree 
\[
\frac{2|E^{\mathcal H}|}{\mathcal V^{\mathcal H}}\ge k-1
\] 
(cf.\ \cite{die00}, p.~98).  
  
  The \emph{diameter} of a connected graph $\mathcal G$ is the number
  $\diam(\mathcal G):=\max\{d^{\mathcal G}(a,b)\mid a,b\in G\}$.
  
  We inductively construct a family $(\mathcal G_i)_{i\ge 1}$ of
  graphs as follows: $\mathcal G_1$ is the graph consisting of two
  vertices and an edge between them. Suppose now that $\mathcal G_i$
  is already defined. Let $\mathcal G_{i+1}'$ be a graph with
  $g(\mathcal G_{i+1}')\ge 2\diam(\mathcal G_i)+1$ and $\chi(\mathcal
  G'_{i+1})\ge 2i+3$. Let $\mathcal G_{i+1}$ be a connected subgraph of
  $\mathcal G_{i+1}'$ with 
\[
\frac{2|E^{\mathcal G_{i+1}}|}{\mathcal
    V^{\mathcal G_{i+1}}}\ge 2i+2. 
\]
Clearly, $g(\mathcal G_{i+1})\ge
  g(\mathcal G'_{i+1})\ge2\diam(\mathcal G_i)+1$.

  Observe that for every $r\ge 1$ and every graph $\mathcal G$, if
  $2r+1< g(\mathcal G)$ then $\ltw^{\mathcal G}(r)\le1$. Moreover,
  if $\mathcal G$ is connected then
  $\ltw^{\mathcal G}(r)= \tw(\mathcal G)$ for all $r\ge\diam(\mathcal
  G)$. For every $i\ge 1$ and $\diam(\mathcal G_i)\le
  r<\diam(\mathcal G_{i+1})$, we let $f(r):=\max\{\tw(\mathcal
  G_i),\ltw^{\mathcal G_{i+1}}(r)\}$. We claim that $\ltw^{\mathcal
  G_i}(r)\le f(r)$ for all $i,r\ge 1$. This is obvious for $i=1$. For
  $i\ge 2$, we have to
  distinguish between three cases: If $r<\diam(\mathcal
  G_{i-1})\le\frac{1}{2}(g(\mathcal G_i)-1)$, then $\ltw^{\mathcal
  G_{i}}(r)\le1\le f(r)$. If $\diam(\mathcal
  G_{i-1})\le r<\diam(\mathcal G_i)$, then $\ltw^{\mathcal
  G_i}(r)\le f(r)$ immediately by the definition of $f$. If
  $r\ge\diam(\mathcal G_i)$, then $\ltw^{\mathcal G_i}(r)=\tw(\mathcal
  G_i)\ge f(r)$.

  Thus the class $\CC:=\{\mathcal G_i\mid i\ge 1\}$ has bounded local
  tree-width. On the other hand, for every $i\ge 2$ we have
$
||\mathcal G_i||\ge|E^{\mathcal G_i}|\ge i|G_i|.
$
\end{example}

\section{The main algorithm}
Throughout this section, we fix a vocabulary $\tau$. We shall first prove two lemmas.

\begin{lemma}\label{lem:ma1}
Let $\CC$ be a class of $\tau$-structures of bounded local tree-width and $r,w\ge
1$. Then there is an algorithm that solves the
following problem in time $O(||\mathcal T||)$:

\prob{Structure $\mathcal A\in \CC$, $(r,w)$-tree cover $\mathcal T$ of $\mathcal
  A$}{Compute $K_T:=\{a\in A\mid N_r^{\mathcal A}(a)\subseteq T\}$ for
  all $T\in\mathcal T$}
\end{lemma}

\proof Observe that $||\mathcal A||=O(||\mathcal T||)$, because by
Lemma \ref{lem:linsize}, for all $T\in\mathcal T$ we have $||\langle
T\rangle^{\mathcal A}||=O(|T|)$.

Without loss of generality we can assume that $\mathcal A$ is a graph;
if not we first compute its Gaifman graph. This is possible in time
$O(||\mathcal A||)$. Furthermore, we can assume that the universe $A$ of
$\mathcal A$ is the set
$\{1,\ldots,n\}$ (see the appendix of \cite{flufrigro00} for details).

Let $T\in\mathcal T$, we show how to compute $K_T$ in time $O(|T|)$.
We suppose that $T$ is given as a list $a_1,\ldots,a_m$ of its
elements. Our algorithm is displayed in Figure \ref{fig:alg3}. $K_T$
is computed iteratively, during the computation the current state of
the set is stored in an array $K$ of length $n$. Note that we do not
initialize the array to 0 in the beginning (we do not have the time to
do that). Instead, we maintain a second ``control array'' $\Gamma$ of
length $m$. The $j$th entry of $\Gamma$ is $a_j$, for $j=1$ to
$m$. $\Gamma$ is initialized to these values in Line 1.
Then at every stage in the computation, the set of all elements
represented by the array $K$ is 
\[
S(K):=\{a\in A\mid K[a]\in\{1,\ldots,m\}\text{ and }\Gamma[K[a]]=a\}.
\]
After Line 2 is executed, we have $S(K)=T$. 

Now the main loop in Lines 3--13 iteratively removes those elements
from $S(K)$ whose neighbors are not all contained in $S(K)$. Thus
after the $i$th iteration we have
\[
S(K)=\{a\in T\mid N_i^{\mathcal A}(a)\not\subseteq T\}.
\]
So once we enter Line 15, we have $S(K)=K_T$. Lines 15--17 retrieve
this set from the array $K$. 
\begin{figure}[ht]
\begin{center}
\fbox{\parbox{12cm}{
\textbf{Input: }${\mathcal A}$,
$T=\{a_1,\ldots,a_m\}\subseteq A$
\begin{algorithm}
\im{0}
\FOR\ $j=1$ \TO\ $m$ \DO\ $\Gamma[j]:=a_j$ \OD
\im{0}
\FOR\ $j=1$ \TO\ $m$ \DO\
$K[a_j]:=j$
\OD
\im{0}
\FOR\ $i=1$ \TO\ $r$ \DO
\im{1}
$\text{temp}:=\emptyset$
\im{1}
\FOR\ $j=1$ \TO\ $m$ \DO
\im{2}
\IF\ $a_j$ has a neighbor $b$ such that
\im{4}
$\Big(\hspace{5mm}K[b]\not\in\{1,\ldots,m\}$
\im{4.2}
\OR\ $\big(K[b]\in\{1,\ldots,m\}$
\AND\ $\Gamma[K[b]]\neq b\big)\Big)$ \THEN 
\im{3}
$\text{temp}:=\text{temp}\cup\{a_j\}$
\im{2}
\FI
\im{1}
\OD
\im{1}
\FORALL\ $a\in\text{temp}$ \DO
\im{2}
$K[a]:=0$
\im{1}
\OD
\im{0}
\OD
\im{0}
$K_T:=\emptyset$
\im{0}
\FOR\ $j=1$ \TO\ $m$ \DO
\im{1}
\IF\ $K[a_j]=j$ \THEN\ $K_T:=K_T\cup\{a_j\}$ \FI
\im{0}
\OD
\end{algorithm}
\textbf{Output: }$K_T$
}}
\end{center}
\caption{}\label{fig:alg3}
\end{figure}

Let us analyze the running time of the algorithm. Lines 1 and 2
require time $O(m)$. To test the condition of Lines 7--8 requires
constant time for each $b$. To test the condition of Lines 6--8, we
have to step through the list of vertices adjacent to $a_j$ until
either we find a $b$ that does not satisfy the condition or we have
checked all neighbors. This requires a constant amount of work for
every edge with one endpoint $a_j$ and the other endpoint in $S(K)$
and an additional constant amount of work in case we find a neighbor
not in $S(K)$. Thus the execution of the loop in lines 5--11 requires
time $O(m+|E^{{\mathcal A}}\cap T^2|)\le O(||\langle
T\rangle^{\mathcal A}||)=O(m)$. The loop in Lines 12--14 also requires
time $O(m)$. Thus every iteration of the main loop requires time
$O(m)$. Since we treat the number $r$ of iterations as constant, the
overall time required by Lines 3--15 is $O(m)$. Since $K_T\subseteq
T$, Lines 16--19 also require time $O(m)$.  \proofend

\begin{lemma}\label{lem:ma2}
Let $\CC$ be a class of structures of bounded local tree-width and $r,m\ge
1$. Then the following problem can be solved in time $O(|A|)$:

\prob{Structure $\mathcal A\in \CC$, set $P\subseteq A$}{Decide if there
  exist $a_1,\ldots,a_m\in P$ such that $d^{\mathcal A}(a_i,a_j)>r$}
\end{lemma}

\proof
Let $f:\mathbb N\rightarrow\mathbb N$ be a function bounding
the local tree-width of the structures in $\CC$.

Let $\mathcal A\in \CC$ and $P\subseteq A$. Our algorithm is displayed
in Figure \ref{fig:alg4}. It proceeds in two phases. 
\begin{figure}[ht]
\begin{center}
\fbox{\parbox{12cm}{
\textbf{Input: }${\mathcal A}\in \CC$, $P\subseteq A$

\begin{algorithm}
\im{0}
$Q:=P$
\im{0}
$l:=0$
\im{0}
\WHILE\ $Q\neq\emptyset$ \AND\ $l<m$ \DO
\im{1}
$l:=l+1$
\im{1}
choose $a_l\in Q$ arbitrarily
\im{1}
$Q:=Q\setminus N_r^{\mathcal A}(a_l)$
\im{0}
\OD
\im{0}
\IF\ $l=m$ \THEN\ 
\im{1}
\ACCEPT
\im{0}
\ELSE\ 
\im{1}
\IF\ $l=0$ \THEN\ \REJECT\ \FI
\im{0}
\FI

\medskip
\im{0}
compute $H:=N_{2r}^{\mathcal A}(\{a_1,\ldots,a_l\})$
\im{0}
\IF\ $\big(\langle H\rangle^{\mathcal A},P\big)\models\exists x_1\ldots\exists
x_m\Big(\bigwedge_{i=1}^m Px_i\wedge\bigwedge_{1\le i<j\le m}d(x_i,x_j)>r\Big)$ \THEN
\im{1}
\ACCEPT
\im{0}
\ELSE
\im{1}
\REJECT
\im{0}
\FI
\end{algorithm}
}}
\end{center}
\caption{}\label{fig:alg4}
\end{figure}

In the first phase (Lines 1--12) it iteratively computes 
elements $a_1,\ldots,a_i\in P$, for some $i\le m$, such that
$d^{\mathcal A}(a_i,a_j)>r$ for $1\le i<j\le l$ and either $l=m$ or
for all $b\in P$ there is an $i\le l$ such that $b\in N_r^{\mathcal
  A}(a_i)$. If $l=m$, the algorithm accepts. If $l=0$, i.e.\
$P=\emptyset$, then it rejects. Otherwise, it goes into the second
phase (Lines 13--18).

When the algorithm enters Line 13, we have $P\subseteq N_{r}^{\mathcal
  A}(\{a_1,\ldots, a_l\})$. Let $\mathcal H:=\langle N_{2r}^{\mathcal
  A}(\{a_1,\ldots,a_l\})$. Then for all $b,b'\in P$ we have
$d^{\mathcal A}(b,b')\le r\iff d^{\mathcal H}(b,b')\le r$, because
  $P\subseteq N_r^{\mathcal A}(\{a_1,\ldots, a_l\})$ and thus every
  path of length at most $r$ between two elements of $P$ must be
  contained in $H$. Thus there
exist $b_1,\ldots,b_m\in P$ such that $d^{\mathcal A}(b_i,b_j)>r$ if,
and only if, there exist $b_1,\ldots,b_m\in P$ such that $d^{\mathcal
  H}(b_i,b_j)>r$, i.e.\ if the condition in Line 14 is satisfied.
Thus the algorithm is correct

To estimate the running time, we note that $||\langle N_{r}^{\mathcal
  A}(a_i)\rangle^{\mathcal A}||=O(|N_r^{\mathcal A}(a_i)|)$, because
$\CC$ is a class of bounded local tree-width. Since we treat $r$ and $m$
as constants, Lines 1--13 require time $O(|A|)$. It is easy to see
that $\tw(\mathcal H)\le \ltw^{\mathcal A}(2lr)\le f(2lr)$. Thus the
condition in Line 14 can also be checked in time $O(|A|)$ by
Courcelle's Theorem \ref{theo:cou}.  
\proofend

We are now ready to prove our main results, Theorems \ref{theo:mt1} and
\ref{theo:mt2}. Recall the statements:

\begin{quote}\itshape
Let $\CC$ be a class of structures of bounded local tree-width and $\phi$ a
sentence of first-order logic. 

\begin{enumerate}
\item
For every $k\ge 1$ there is an algorithm that decides whether a given structure
  $\mathcal A\in \CC$ satisfies $\phi$ in time $O(|A|^{1+(1/k)})$.
\item
If $\CC$ is locally tree-decomposable, then there is an algorithm that
solves the problem in time $O(|A|)$.
\end{enumerate}
\end{quote}

\proof
We describe the algorithm for (1) and then explain how it has to be
modified to obtain (2).

By Gaifman's Theorem \ref{gaifman}, without loss of generality we can
assume that $\phi$ is of the form
\[
\exists x_1\ldots\exists x_m\big(\bigwedge_{1\le i<j\le
  m}d(x_i,x_j)>2r\wedge \bigwedge_{1\le i\le
  m}\psi(x_i)\big),
\]
for suitably chosen $r,m\ge1$ and an $r$-local $\psi$.

Let $k\ge 1$ and $f:\mathbb N\to\mathbb N$ be a function bounding the local
tree-width of the structures in $\CC$. Let $\tau$ be the vocabulary of
the sentence $\phi$; without loss of generality we can assume that all
structures in $\CC$ are $\tau$-structures.

Figure \ref{fig:alg2} shows our algorithm.
\begin{figure}[ht]
\begin{center}
\fbox{\parbox{12cm}{
\textbf{Input:} Structure $\mathcal A\in \CC$

\begin{algorithm}
\im{0}
compute an $(r,2kr)$-neighborhood cover $\mathcal N$ of $\mathcal A$
of size $O(A^{1+(1/k)})$
\im{0}
\FORALL\ $N\in\mathcal N$ \DO
\im{1}
compute $K_N:=\{a\in N\mid N_r^{\mathcal A}(a)\subseteq N\}$
\im{0}
\OD
\im{0}
\FORALL\ $N\in\mathcal N$ \DO
\im{1}
compute $P_N:=\big\{a\in K_N\bigmid \langle N\rangle^{\mathcal
  A}\models\psi(a)\big\}$.
\im{0}
\OD
\im{0}
compute $P:=\bigcup_{N\in\mathcal N}P_N$
\im{0}
\IF\ there are $a_1,\ldots,a_m\in P$ such that $d(a_i,a_j)>2r$ for
$1\le i<j\le k$ \THEN\
\im{1}
\ACCEPT
\im{0}
\ELSE
\im{1}
\REJECT
\im{0}
\FI 
\end{algorithm}
}}
\end{center}
\caption{}\label{fig:alg2}
\end{figure}

To see that the algorithm is correct, note that since $\psi(x)$ is
$r$-local we have $P=\{a\in A\mid \mathcal A\models\psi(a)\}$.

\medskip
So we shall prove that the algorithm can be implemented as an
$O(n^{1+(1/k)})$-algorithm, where $n:=|A|$ is the cardinality of the
input structure. 

Line 1 requires time $O(n^{1+(1/k)})$ by Corollary \ref{cor:ncov}.
Lines 2--4 require time $O(||\mathcal N||)$ by Lemma \ref{lem:ma1}. For
every $N\in\mathcal N$, Line 6 requires time $O(|N|)$ by Lemma
\ref{lem:cou}. Thus the loop in Lines 5--7 also requires time $O(||\mathcal
N||)$. Clearly, Line 8 can be performed in time $O(||\mathcal N||)$,
and the condition in Line 9 can be checked in time $O(|A|)$ by Lemma
\ref{lem:ma2}. Thus the overall running time is $O(||\mathcal
N||)=O(n^{1+(1/k)})$.

\medskip
It remains to prove (2), but this is very easy now. Instead of a
neighborhood cover, in Line 1 of the algorithm we compute tree cover
of linear size. This can be done in linear time by the definition of a
locally tree-decomposable class of graphs. Since the running time of
the rest of the algorithm is linear in the size of the cover, we
obtain a linear time algorithm.
\proofend

\section{Concluding remarks}

\subsection*{Uniformity}  A close look at our
proofs shows that actually for each locally tree-decomposable class
$\CC$ of structures there is a recursive function $f:\mathbb
N\to\mathbb N$ and an algorithm that decides, given a
first-order sentence $\phi$ and a structure $\mathcal A\in \CC$,
whether $\mathcal A\models\phi$ in time $O(f(||\phi||)|A|)$,
where $||\phi||$ denotes the length of the
sentence $\phi$. We can obtain an anlogous uniform version of Theorem
\ref{theo:mt2}.

We stated and proved non-uniform versions of the theorems for the sake
of a clearer presentation.

\subsection*{Dependence on the formula size}
Our algorithm heavily depends on the size of the formula $\phi$,
roughly the hidden multiplicative constant is $k$-fold
exponential in the length of $\phi$, where $k$ is the number of
quantifier-alternations in $\phi$. 

\subsection*{Practical Considerations}
The large hidden constants seem to make our algorithms useless for
practical purposes. Nevertheless, let us briefly discuss a few more practical
aspects.

The main factor contributing to the large constants is the complexity
of the formulas, in particular the number of quantifier alternations.
However, if we think of a database application, we will usually only
have to handle very simple formulas. As matter of fact, most database
queries are so called \emph{conjunctive queries}; they can be
defined by first-order formulas of the form $\exists x_1\ldots\exists
x_k(\alpha_1\wedge\ldots\wedge\alpha_m)$. Such formulas do not have
any quantifier alternation. Moreover, when handling conjunctive
queries we can avoid the second very costly part hidden in our
algorithms, namely the transformation of a formula according to Gaifman's
theorem. For all we know, such a transformation may blow up the
formula size by a non-elementary factor. For conjunctive queries, we
can avoid Gaifman's theorem and instead use algorithmic techniques
similar to those in \cite{epp99a}. With these techniques, the
dependence on the formula size can be reduced to a singly exponential
factor, which seems acceptable because usually in practice we have to
evaluate small formulas (queries) in large structures (databases). The
third costly factor is to compute tree-decompositions. Bodlaender's
linear time algorithm is only of theoretical interest due to
very large hidden constants. More promising seems to be an algorithm
due to Reed \cite{ree92} (improving an earlier algorithm due to Robertson
and Seymour \cite{GMXIII}). For an input graph of tree-width $w$ and
size $n$, this algorithm only computes a
tree-decomposition of width at most $4w$. Its running time is not
linear in $n$ (as Bodlaenders), but $O(n\log(n))$. However, the
algorithm is simple and the hidden constants, though exponentially
depending on $w$, do not seem too large for small values of $w$.
Let us also remark that there are much more efficient algorithms for
computing small width tree-decompositions of planar graphs of small
radius \cite{epp99}.

Nevertheless, as they stand our results are mostly
theoretical. Similarly to Courcelle's Theorem \cite{cou90}, their main
benefit is to provide a quick and simple way to recognize a
property as being linear time computable on certain classes of graphs.
Analyzing the combinatorics of the specific property then, one may
also find a practical algorithm. 

\subsection*{Further Research}
Although Example \ref{ex:bltwsize} shows that for classes $\CC$ of
bounded local tree-width we cannot expect an
algorithm deciding a first-order property of structures $\mathcal
A\in\CC$ in time $O(|A|)$, it does not rule out an $O(||\mathcal
A||)$-algorithm. To obtain such an algorithm, it would be sufficient
to find, for every $r\ge 1$, a $w\ge 1$ and an $O(||\mathcal
A||)$-algorithm that computes an $(r,w)$-tree cover of a structure
$\mathcal A\in\CC$.

As we mentioned, one of the main factors contributing to the heavy
dependence of the running time of our algorithms on the size of the
formula is the transformation into a ``local formula'' according to
Gaifman's theorem. Though this transformation is clearly effective, as
far as we know its complexity has not been studied. We do expect this
complexity to be non-elementary, but this does not rule out the
existence of more efficient algorithms for particular classes of
formulas (such as existential formulas) or the existence of good
heuristics.

In general, we consider it as one of the main challenges for further
research to reduce the dependence on the formula size (not only in our
results, but also in Courcelle's theorem). For example, is
there an algorithm that decides, given a first-order sentence $\phi$
and a planar graph $\mathcal G$ (or a tree, or just a word), whether
$\mathcal G\models\phi$ in time $O(2^{|\phi|}n^c)$ for some
fixed-constant $c$?


\end{document}